\documentclass[reprint,amsmath,nofootinbib,showkeys,fleqn,amssymb,aps,prd, superscriptaddress]{revtex4-2}

\usepackage{newtxtext,newtxmath}
\usepackage[T1]{fontenc}
\usepackage[utf8]{inputenc}
\DeclareRobustCommand{\VAN}[3]{#2}
\let\VANthebibliography\thebibliography
\def\thebibliography{\DeclareRobustCommand{\VAN}[3]{##3}\VANthebibliography}

\usepackage{graphicx}	% Including figure files
\usepackage{amsmath}	% Advanced maths commands
\usepackage{orcidlink}
\usepackage{xcolor}
\definecolor{DarkGreen}{rgb}{0,0.6,0}

\newcommand{{\qsub}}{\bf{q}}
\newcommand{{\rsub}}{{\bf{r}}}
\newcommand{{\xlight}}{{\bf{n}}_{\rm{light}}}
\newcommand{{\xlens}}{{\bf{n}}_{\rm{mac}}}
\newcommand{{\data}}{\bf{D}}
\newcommand{{\datan}}{{\bf{d}}_{\rm{n}}}
\newcommand{{\datanprime}}{{{\bf{d}}_{\rm{n}}^{\prime}}}
\newcommand{{\dimg}}{{\bf{d}}_{\mathcal{I}}}
\newcommand{{\dimgprime}}{{\bf{d}}_{\mathcal{I}}^{\prime}}
\newcommand{{\dptsrc}}{{\bf{d}}_{\rm{ptsrc}}}
\newcommand{{\dptsrcprime}}{{\bf{d}}_{\rm{ptsrc}}^{\prime}}
\newcommand{{\dfr}}{{\bf{d}}_{\rm{fr}}}
\newcommand{{\dfrprime}}{{\bf{d}}_{\rm{fr}}^{\prime}}
\newcommand{{\zlens}}{{z_{\rm{d}}}}
\newcommand{{\zsrc}}{{z_{\rm{s}}}}

\def\mlow{m_{\mathrm{low}}}

\begin{document}

\title{Measurement of the minimum cold dark matter halo mass with strong gravitational lensing.}

\author{A.~M.~Nierenberg}%$^\dagger$}
\email{anierenberg@ucmerced.edu}
\affiliation{University of California, Merced, 5200 N Lake Road, Merced, CA 95341, USA}

\author{D.~Gilman}
\affiliation{Department of Astronomy $\&$ Astrophysics, University of Chicago, Chicago, IL 60637, USA}
\affiliation{Brinson Prize Fellow}

\author{T.~Treu\orcidlink{0000-0002-8460-0390}}
\affiliation{Department of Physics and Astronomy, University of California, Los Angeles, CA,  90095, USA}

\author{X.~Du\orcidlink{0000-0003-0728-2533}}
\affiliation{Department of Physics and Astronomy, University of California, Los Angeles, CA,  90095, USA}
\affiliation{School of Physics, Nankai University, Tianjin 300071, P. R. China}

\author{C.~Gannon\orcidlink{0009-0009-0443-3181}}
\affiliation{University of California, Merced, 5200 N Lake Road, Merced, CA 95341, USA}

\author{H.~Paugnat\orcidlink{0000-0002-2603-6031}}
\affiliation{Department of Physics and Astronomy, University of California, Los Angeles, CA,  90095, USA}

\author{S.~Birrer\orcidlink{0000-0003-3195-5507}}
\affiliation{Department of Physics and Astronomy, Stony Brook University, Stony Brook, NY 11794, USA}

\author{A.~J.~Benson\orcidlink{0000-0001-5501-6008}}
\affiliation{Carnegie Institution for Science, Pasadena CA 91101, USA }

\author{K.~N.~Abazajian\orcidlink{0000-0001-9919-6362}}
\affiliation{Department of Physics and Astronomy, University of California, Irvine, CA 92697-4575, USA}
    
\author{T.~Anguita\orcidlink{0000-0003-0930-5815}}
\affiliation{Instituto de Astrofisica, Departamento de Fisica y Astronomia, Facultad de Ciencias Exactas, Universidad Andres Bello, Chile}

\author{S.~G.~Djorgovski\orcidlink{0000-0002-0603-3087}}
\affiliation{California Institute of Technology, Pasadena CA 91125, USA }
    
\author{S.~F.~Hoenig\orcidlink{0000-0003-3030-2360}}
\affiliation{School of Physics and Astronomy, University of Southampton, Southampton SO17 1BJ, United Kingdom }

\author{R. E. Keeley\orcidlink{0000-0002-0862-8789}}
\affiliation{Department of Physics and Astronomy, University of California, Irvine, CA 92697-4575, USA}

\author{A.~Kusenko\orcidlink{0000-0002-8619-1260}}
\affiliation{Department of Physics and Astronomy, University of California, Los Angeles, CA,  90095, USA}
\affiliation{Kavli Institute for the Physics and Mathematics of the Universe (WPI), UTIAS, The University of Tokyo, Kashiwa, Chiba 277-8583, Japan}

\author{H.~R.~Larsson\orcidlink{0000-0002-9417-1518}}
\affiliation{University of California, Merced, 5200 N Lake Road, Merced, CA 95341, USA}

\author{L.~A.~Moustakas}
\affiliation{Jet Propulsion Laboratory, California Institute of Technology, 4800 Oak Grove Dr, Pasadena, CA 91109}

\author{P. ~Mozumdar \orcidlink{0000-0002-8593-7243}}
\affiliation{Department of Physics and Astronomy, University of California, Los Angeles, CA, 90095, USA}

\author{W.~Sheu\orcidlink{0000-0003-1889-0227}}
\affiliation{Department of Physics and Astronomy, University of California, Los Angeles, CA,  90095, USA}
    
\author{D.~Sluse\orcidlink{}}
\affiliation{STAR Institute, University of Li\`ege, Quartier Agora - All\'ee du six Ao\^ut, 19c B-4000 Li\`ege, Belgium }
    
\author{D.~Stern\orcidlink{}}
\affiliation{Jet Propulsion Laboratory, California Institute of Technology, 4800 Oak Grove Dr, Pasadena, CA 91109}

\author{D.~Williams\orcidlink{0000-0002-8386-0051}}
\affiliation{Department of Physics and Astronomy, University of California, Los Angeles, CA,  90095, USA}

\author{K.~C.~Wong\orcidlink{0000-0002-8459-7793}}
\affiliation{Research Center for the Early Universe, Graduate School of Science, The University of Tokyo, 7-3-1 Hongo, Bunkyo-ku, Tokyo 113-0033, Japan}

\begin{abstract}
We explore the lowest mass limit that can be placed on the halo mass function in CDM using 28 strong gravitational lenses. For this purpose, we study an extreme model in which the halo mass function and mass-concentration relation follow CDM, with a sharp cutoff at some mass scale, $\mlow$. 
Lensing provides a unique window into this quantity as it does not depend on the presence of  baryons in dark matter halos and also allows the detection of low mass halos at cosmological distances, both in the lens galaxies and along the line-of-sight. Our model incorporates the effects of tidal stripping of subhalos, leading to the presence of many subhalos below a given model cutoff scale. We place an upper limit on the low-mass cutoff of the halo mass function of $\mlow<10^{8.3}$  M$_\odot$ at 10:1 odds using a prior for the normalization of the subhalo mass function from the semi-analytic model {\tt galacticus} and $\mlow<10^{8.2}$  M$_\odot$ at 10:1 odds using a prior from $N$-body simulations.  These limits are comparable to, or stronger than, existing constraints based on Milky Way satellite galaxies. Based on these results, we forecast more than an order of magnitude improvement with a sample of 200 quadruply imaged quasar lenses. This number represents a small subset of the thousands that are anticipated to be discovered by Rubin, Euclid, and Roman. Furthermore, with this larger sample of lenses we expect to directly constrain the normalization of the subhalo mass function, thereby eliminating a major source of uncertainty in our current measurements.

\end{abstract}


\maketitle

\textit{Introduction}---Early high-resolution dark matter $N$-Body simulations of Cold Dark Matter (CDM) structure made a key prediction: that there must be a significant number of completely dark subhalos around the Milky Way \citep{diemandEarthmassDarkmatterHaloes2005, Moore_subhalos_1999, Kauffman_subhalos_93, klypin_missing_sats_99}. This was based on the CDM model prediction that there should be thousands of subhalos around the Milky Way, in comparison with only tens of known Milky Way satellites.

Since that time, the number of known low-mass satellite galaxies in the Local Group has dramatically increased thanks to modern all-sky surveys which have discovered satellites down to stellar masses of $\sim$ 1000 $M_\odot$ (see e.g. \citep{Doliva-Dolinsky_LG_sats_25} and references therein for a review). Although direct mass estimates of the dark matter halos of these objects are highly uncertain, as these satellites contain few stars for which kinematics can be measured, and the stars occupy only the central tens of parsecs of the dark matter halos \citep{Battaglia_sat_kinematics_2022},
they can still be used to constrain the properties of dark matter through abundance matching. Abundance matching studies compare the predicted number of dark matter halos from high-resolution simulations to the known satellites of the Milky Way to place limits on the scale at which dark matter halos must exist around our Galaxy, thereby bypassing the need for a direct measurement of the subhalo mass function. An accurate estimate of the completeness of observations of satellite galaxies is crucial to these measurements \citep{Tollerud_mw_lf_2008, kim_satcompleteness_2018, nadler_milky_2020}. \citet{nadler_milky_2020} used abundance matching to place an upper limit on the lowest mass halos that must exist, prior to the effects of tidal stripping, of $\sim$10$^{8.4}$ M$_\odot$.

%An important caveat is that such measurements are made of Milky Way satellites, as this is the only region close enough to detect the ultra-faint galaxies that occupy these halos. Outside of the local volume, such measurements are extremely difficult.

Gravitational lensing provides a means of extending the measurement of the halo mass function to cosmological volumes, as it relies on the total mass of objects and does not require the detection of luminous structure \citep{dalal_direct_2002, mao_evidence_1998, vegetti_strong_2023}. Recent measurements, using a combination of quasar mid-IR and narrow-line emission flux ratios, have placed some of the strongest limits to date on a Warm Dark Matter (WDM) turnover in the halo mass function \citep{JWST_4_Gilman}. In this work, we seek to address a different question: the extent to which gravitational lenses can place limits on the lowest mass halos. We adopt an empirical model in which we assume the CDM prediction for the mass-concentration relation and mass function down to a sharp cutoff at some lower mass limit at all redshifts. This model is not tied to any realistic particle model of dark matter, which would, among other things, predict an evolution of the low mass limit with redshift, and does not reproduce the CDM mass-concentration relation. Instead, the model enables us to determine in a robust and conservative  manner the minimum mass scale at which current strong lensing measurements
can confirm a fundamental prediction of CDM, namely that a population of low-mass halos devoid of luminous matter exists throughout the Universe.

%Importantly, the halo mass limit derived here corresponds to a lower bound on the particle dark matter mass. In the simplest models, dark matter particles are produced thermally and retain their thermal momentum distribution during early structure formation, which uniquely determines their kinematic behavior \cite{Colombi:1995ze}. The only remaining parameter governing the linear dynamics of thermal dark matter is therefore its particle mass. For thermal dark matter with particle masses of order $\sim$1 to $\sim$10 keV, free streaming suppresses the matter power spectrum on small scales, reducing the abundance of the lowest-mass field halos near and below the mass enclosed within the free-streaming horizon. In cases where the dark matter has a nonthermal momentum distribution, the mapping between particle mass and halo mass suppression scale is no longer uniquely determined. This degeneracy has been explored extensively for oscillation-produced sterile neutrino dark matter; however, the suppression scale can vary significantly depending on the production mechanism (e.g., \citet{Abazajian:2019ejt}). Since subhalo mass limits are more constraining of particle properties \cite{JWST_4_Gilman}, a detailed mapping between the halo mass limit and particle dark matter mass is deferred to future work. 

%prove the existence of CDM-like halos.

\textit{Strong lensing dark matter detection}---
In a strong gravitational lens, the image positions and relative image magnifications are determined by the gravitational potential. Small perturbations to the main lens mass distribution in the form of low-mass structure in the lens and along the line of sight can significantly alter the relative image magnifications relative to a smooth mass expectation while leaving the positions unchanged. Two decades ago, several works noted that the sensitivity of strong lenses with unresolved sources to low-mass halos could be used to study dark matter substructure \citep{dalal_direct_2002,mao_evidence_1998}. With the advent of new instruments enabling spatially resolved spectroscopy \citep{nierenberg_detection_2014,nierenberg_probing_2017,nierenberg_double_2020} and measurements of quasar warm dust emission \citep{Nierenberg_0405_2023, Keeley_2024,JWST_III} the sample of lenses to which we can apply this method has increased by a factor of three. 

\textit{Data}---
We study the 28 lenses from the James Webb Lensed Quasar Dark Matter Survey \citep{Nierenberg_0405_2023}. This sample was selected from all quadruply imaged quasars with measured microlensing-free flux ratios, which do not contain a significant disk deflector. These include 2 lenses with narrow-line emission \citep{nierenberg_detection_2014, nierenberg_double_2020}, and 26 lenses with warm dust flux ratios measured with JWST as part of program GO-2046 (PI Nierenberg) \citep{Nierenberg_0405_2023, JWST_III}. While the flux ratios provide a highly sensitive probe of small scale perturbations due to low-mass halos, imaging of the quasar host galaxy can significantly improve the constraining power by reducing uncertainty on the large-scale mass distribution of the deflector, as demonstrated by \citet{2024GilmanTurbo, JWST_4_Gilman}. In this work, we use the same imaging data set and statistical approach described in detail in \citet{JWST_4_Gilman} to incorporate information from the imaging data into the inference of dark matter parameters.  We summarize key aspects of the method here.

\textit{Macromodel}---
We model the large-scale, mass distribution of the lens, or macromodel following the current state of the art in the field, which is a power-law ellipsoid mass distribution with $m=1,3,4$ multipoles, and external shear. When detected, we also include luminous companion galaxies close in proximity to the lensed images modeled as Singular Isothermal Spheres. Priors on lens model parameters were selected based on the known properties of strong lenses from \citep{Auger_slacs_fdm} and massive elliptical galaxies \citep{hao_isophotal_2006, Oh_improving_2024} as outlined in detail by \citet{JWST_4_Gilman}.  Traditionally, only the image positions have been used to constrain the mass distribution of the deflector. We follow \citet{JWST_4_Gilman} in also incorporating lensed arc information, when available, to constrain the macromodel, as described below. 

\textit{Dark matter model}--- We investigate the joint signal of dark matter halos in the lens (subhalos) and along the line of sight (field halos).  Field halos are drawn from a modified Sheth-Tormen mass function \cite{Sheth_Torman_2001}:

\begin{equation}
		\label{eqn:mfunclos}
		\frac{\mathrm{d}^2 N}{\mathrm{d}m \mathrm{d}V} = \delta_{\rm{LOS}} \left[1+\xi\left(m_{\rm{host}},z_{\rm{d}}\right)\right] \frac{\mathrm{d}^2 N_{\rm{ST}}}{\mathrm{d}m \mathrm{d}V}.
	\end{equation}

\noindent Where $N$ is the model predicted number of halos, $N_{\rm{ST}}$ is the number of halos predicted by the Sheth-Torman mass function, while $\xi$ accounts for correlated structure following  \citet{Lazar++21}. In addition, we allow for potentially over- or under-dense sightlines via the dark matter model parameter $\delta_{\rm{LOS}}$, which is sampled between 0.9 and 1.1, with 1 being the Sheth-Tormen prediction.

Halos are drawn from this mass function between a lower mass limit of $\mlow$, which is a sampled dark matter model parameter and a fixed upper limit of $10^{10.7}$ M$_\odot$.  The upper limit of the mass function corresponds to the halo mass at which we expect to detect luminous satellites embedded in the dark matter halos, given the observation depth. Such halos are explicitly included in our analysis with steeper, Singular Isothermal Sphere density profiles to account for the effects of baryons.  
The lower limit, $\mlow$, is the parameter under investigation in this study. It represents an absolute cutoff in the halo mass function at all redshifts. This is not motivated by realistic particle physics or structure formation model but instead provides a limiting estimate of the sensitivity of the data to the presence low mass halos.  We allow this to vary between $6<\log_{10}[\mlow/\rm{M}_\odot]<10$. The lower value of the allowed range is our estimated sensitivity floor to for a population of halos to produce a measurable flux ratio anomaly given the current number of lenses, lens modeling uncertainties and flux measurement uncertainties.

Subhalos are halos which have been accreted into the larger virial radius of the main lens dark matter halo. The subhalo mass function at infall (prior to the effects of tidal stripping) is modeled as: 

\begin{equation}
		\label{eqn:subhalomfunc}
		\frac{\mathrm{d}^2 N}{\mathrm{d}m \mathrm{d}A} = \frac{\Sigma_{\rm{sub}}}{m_0}\left(\frac{m}{m_0}\right)^{-\alpha} \mathcal{F}\left(m_{\rm{host}},z\right).
	\end{equation}

 The logarithmic slope of the subhalo mass function, $\alpha$, is varied between 0.9--1.1, which encompasses both the prediction from pure dark matter simulations as well as the possible impact of baryons
\citep{Giocoli++08, Springel++08, fiacconi_cold_2016, Benson20, Gannon_2025}.
$\Sigma_{\rm{sub}}$ is the normalization of the subhalo mass function at 10$^8$ M$_\odot$ for a 10$^{13}$ M$_\odot$ host halo at redshift 0.5 and $\mathcal{F}\left(m_{\rm{host}},z\right)$ represents the scaling of the subhalo mass function with host halo mass and redshift:
\begin{equation}
		\label{eqn:shmfscaling}
		\mathcal{F}\left(m_{\rm{host}},z\right) = \left(m_{\rm{host}}/10^{13} \mathrm{M}_{\odot}\right)^{k_1}\left(z+0.5\right)^{k_2},
	\end{equation}
	with $k_1=0.55$ and $k_2=0.37$ measured from \citep{Gannon_2025}.

The logarithm of the host halo mass is sampled over a Gaussian prior with $\log_{10}[m_{\rm{host}}/\rm{M}_{\odot}$] having a mean of 13.3 and standard deviation of 0.3
based on the measurement of halo mass distribution for a sample of representative galaxy-scale strong lenses by \citet{Lagattuta_2010} .

Subhalos undergo significant tidal interactions as they are accreted into the main lensing halo. We use the analytic framework developed by \citet{Du++25} to statistically model the subhalo tidal evolution as a function of key parameters, including infall time and subhalo concentration.  As with other state of the art subhalo evolution models, this model predicts that typical subhalos in the projected $\sim$ 5 arcseconds of strong lens-like halos have lost about 95\% of their mass since infall. Owing to these effects, our forward modeling simulations contain many subhalos with final masses significantly lower than the field halo cutoff of $\mlow$. 

Current $N$-body simulations and semi-analytic models differ in their predictions for the amplitude of the subhalo mass function by a factor of two \citep{Gannon_2025}. To account for this theoretical uncertainty, we explore two different priors on the projected normalization of the subhalo mass function, $\Sigma_{\rm {sub}}$, one reflecting the prediction from $N$-body simulations and one from the semi-analytic model {\tt galacticus} \cite{benson_g_2012, Shengqi_orbital_galacticus, Du++24}.  

Field halo and subhalo concentrations are determined from the relations in \citet{Diemer_2019_CDM_concentration}. Subhalo concentrations are evaluated at the infall redshift.
We note that our forward modeling includes halos with masses several orders of magnitude below where the mass-concentration relation was calibrated ($\sim 10^{10}$ M$_\odot$) in that study. Reassuringly, this model produces consistent results for the mass-concentration relation down to masses of $\sim10^6$ M$_\odot$ measured in ultra high resolution zoom in simulations by \citet{Wang_2020_mc}. As we describe below, significant changes to halo concentrations can have an important impact on the predicted lensing signal for a fixed halo mass function.

	\begin{table*}
		\setlength{\tabcolsep}{12pt}
		\caption{\label{tab:tableqsub} Description of the dark matter hyper-parameters. }
		\begin{tabular}{cccc}
			\hline
			Hyper-parameter & Description & Sampling distribution &  Remarks\\	
			\hline
			$\delta_{\rm{LOS}}$ & rescales the amplitude of the& $\mathcal{U}\left(0.9, 1.1\right)$ & $\delta_{\rm{LOS}}=1$ corresponds to the \\
			& field halo mass function  & & Sheth--Tormen prediction\\ \\ 
			$\alpha$ & logarithmic slope of the & $\mathcal{U}\left(-1.95, -1.85\right)$ & CDM predicts $\alpha \sim -1.9$\\
			& subhalo mass function at infall & & \\ \\ 
			$\Sigma_{\rm{sub}} \ \left[\rm{kpc^{-2}}\right]$ & amplitude of the differential & $\log_{10} \mathcal{U}\left(-2.2, 0.2\right)$ & $N$-body predicts $\sim 0.1 \ \rm{kpc^{-2}}$\\
			& subhalo mass function at infall & & {\tt galacticus} predicts $\sim 0.15 \ \rm{kpc^{-2}}$ \\
            &  & &  \\  
			$m_{\rm{low}} \left[\mathrm{M}_{\odot}\right]$ & field halo cutoff & $\log_{10} \mathcal{U}\left(6.0, 10\right)$ &  \\
			\hline
		\end{tabular}
	\end{table*}

\textit{Dark matter forward modeling}---
The details of the statistical model we use are laid out in \citet{JWST_4_Gilman}. We use a forward modeling approach to draw realizations of dark matter halos from the dark matter model parameters (Table \ref{tab:tableqsub}) as well as to sample over the lens mass distribution (macromodel parameters) described previously. For each realization of dark matter and macromodel parameters, we compute the model-predicted image magnifications, which are compared to observed magnifications to compute a likelihood for that set of parameters. We use {\tt lenstronomy} \citep{birrer_lenstronomy_2018, birrer_lenstronomy_2021} to perform gravitational lensing calculations and to optimize free lens parameters to match the image positions for each dark matter realization. A detailed description of how the lensing calculation is computed to ensure matching to image positions is provided by \citet{JWST_4_Gilman}. We generate between 0.5--20 million samples per lens to ensure convergence.

We also incorporate the extended arcs from the lensed quasar host galaxies, when detected. This information was included via a separate suite of simulations. In these simulations, we generate realizations from the dark matter and macromodel parameters in an identical way to the previous step and require that the image positions be matched. The difference in this step was that for every realization, we perform full ray tracing across the image plane to simultaneously reconstruct the lens mass profile and the extended arcs. We then compute the likelihood using all pixels in the image plane from the model predicted and observed imaging data. Importantly, in this step, the likelihood does not include the flux ratios. We repeat these calculations for 200,000 different realizations of dark substructure to obtain importance sampling weights on the macromodel parameters marginalized over many possible realizations. 
%The recovered simulated images was compared to the real image to compute an imaging data likelihood. We marginalized over the dark matter parameters from all of the realizations (Table \ref{tab:tableqsub}) leaving behind a measurement of the overall likelihood of the macromodel parameters across all dark matter realizations. These likelihoods are independent of the flux ratios and rely only on the goodness of fit to the imaging data and lensed quasar positions.
%The likelihoods are then applied as weights to the full forward modeling simulations calculated in the first steps, so that forward model iterations with macromodel parameters favored by the imaging only analysis are given preferential weight in the final analysis. 
%By decoupling the imaging analysis from the flux ratio computation in the first step, we are able to efficiently sample over many sets of macromodel parameters. We find that the macromodel parameter weights are well estimated with about two hundred thousand realizations per lens.

In the final step of the statistical inference, we apply the importance weights computed from the extended arcs to the lens models generated for the dark matter inference, for which we compute the flux ratios. This ensures that each realization is weighted by the likelihood of both the flux ratios and the macromodel parameters.  The lensed arcs provide a significant reduction in the uncertainty of the model predicted flux ratios, primarily through a reduction in the uncertainty in the deflector ellipticity and orientation and to a lesser extend the parameters of the multipoles relative to the initial prior \citep[see e.g. Figs. 9--17 of][]{JWST_4_Gilman}.

The statistical methods used in this work have been validated on mock data and have been shown to accurately recover properties of the halo mass function for a variety of dark matter scenarios \citep{2024GilmanTurbo} in the presence of realistic, complex macromodels.

\begin{figure}
    \includegraphics[trim=1cm 1cm 1cm
    0.5cm,width=0.45\textwidth]{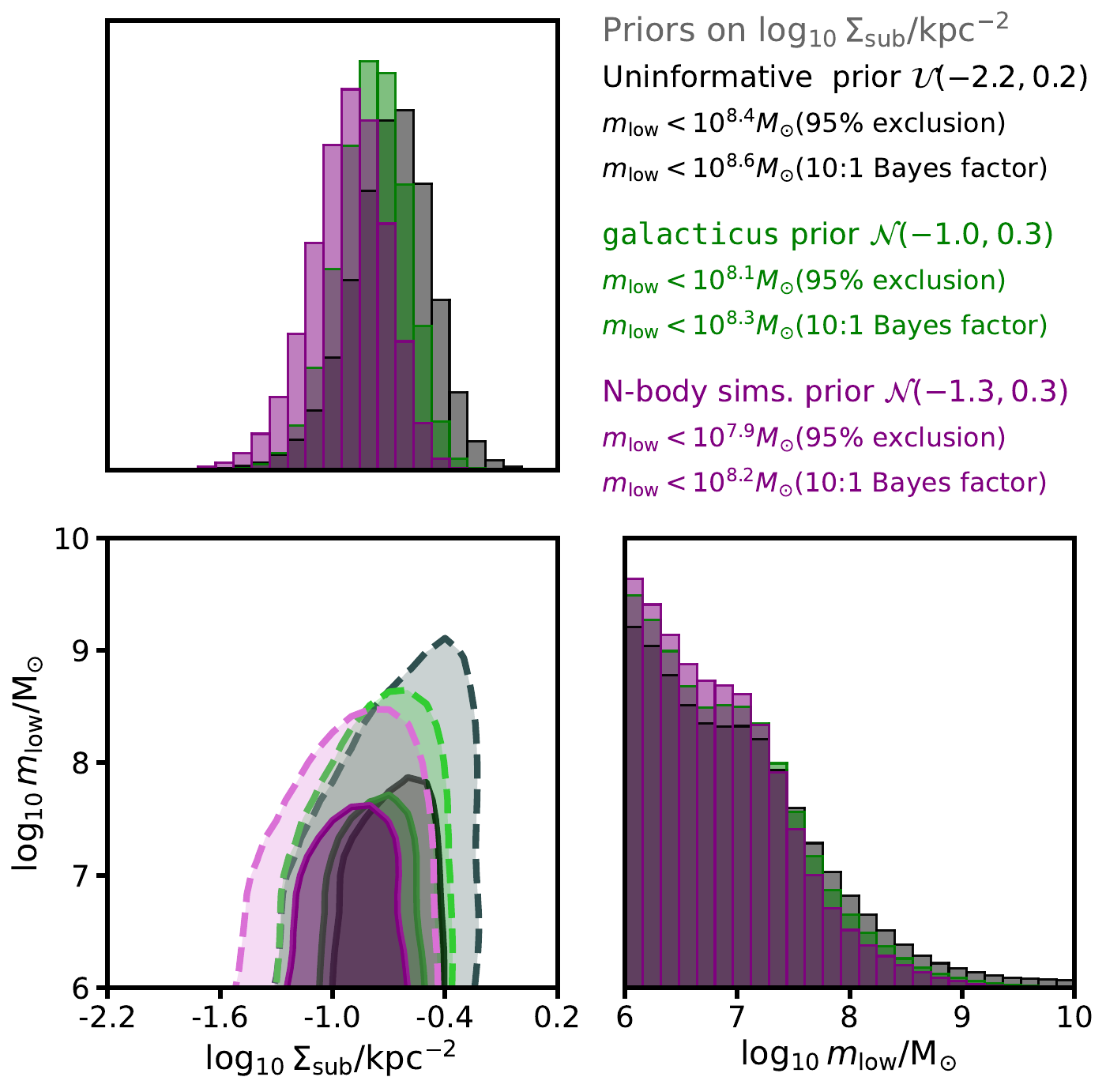}
    \caption{\label{fig:results} The joint posterior probability distribution for a cutoff in the halo mass function ($\mlow$), and normalization of the projected subhalo mass function ($\Sigma_{\rm{sub}}$), shown for three different priors on $\Sigma_{\rm{sub}}$, jointly inferred from 28 lenses. One-sided $95\%$ exclusion limits, and 10:1 Bayes factor constraints taken with respect to the peak of the marginal distribution, are summarized in the top right. Contours show $68\%$ and $95\%$ confidence regions.}
\end{figure}

\textit{Results}---
Figure~\ref{fig:results} shows the posterior probability distribution for the dark matter parameters $\Sigma_{\rm{sub}}$, and $\mlow$, marginalizing over $\alpha$ and $\delta_{\rm{los}}$ which were not well constrained. We show the result for three choices of prior for the normalization of the subhalo mass function.  

The most agnostic prior is a uniform prior for $\log_{10}\Sigma_{\rm{sub}}$ between $-2.2$ and 0.2. This prior extends approximately a factor of 10 above and below the predictions from both $N$-body simulations and {\tt galacticus}. For this prior we measure $\mlow<10^{8.6}$ M$_\odot$ at 10:1 Bayesian odds, and $\mlow<10^{8.4}$ M$_\odot$ at 95\% confidence. Adopting more informative priors, we infer $\mlow<10^{8.3}$ ($10^{8.1}$) M$_\odot$ at 10:1 odds (95\% confidence) using the prior for $\Sigma_{\rm{sub}}$ based on {\tt galacticus} and $\mlow<10^{8.2}$ ($10^{7.9}$) M$_\odot$ at 10:1 odds (95\% confidence) using the prior from $N$-body simulations.

These are among the strongest limits measured to date on the minimum mass scale at which dark matter halos. They are comparable to existing limits from Milky Way satellite galaxies by \citet{nadler_milky_2020} who report $\mlow<10^{8.8}$ at 95\% confidence, using spectroscopically confirmed Milky Way satellites. Including satellites that are not spectroscopically confirmed strengthens the constraint to $\mlow<10^{8.4}$ $M_\odot$ at 95\% confidence. 

Direct comparison between these two studies posses some complexity. First, the prior ranges on $\mlow$ used in the two analyses are different, therefore the 95\% confidence intervals are not directly comparable.  Second, limits from \citet{nadler_milky_2020} are based exclusively on subhalos and are relative to the peak mass that subhalo ever reached. Given that subhalos lose mass over an extended period of time prior to infall, the connection between mass definitions between the mass at infall and the peak infall mass can differ by up to 40\% \cite{Ahvazi_infall_vs_peak}. To some extent this uncertainty is captured by our two priors on the normalization of the subhalo mass function. The {\tt galacticus} model does not account for pre-infall tidal stripping while the $N$-body prior does. These differences are also mitigated by the fact that our measurement probes both line-of-sight halos as well as subhalos. In summary, we do not aim to directly compare the stringency of these two measurements, only to demonstrate that both gravitational lensing and Milky Way satellite counts measurements indicate the existence of a significant population of dark matter halos at masses of 10$^{8.5}$ M$_\odot$.

%A recent gravitational imaging detection by \cite{Vegetti_challenge} favors the presence of a low-mass dark object with mass of $\sim5\times10^{6}$ M$_\odot$. This object compact, with best fitting models containing a point-mass which contains a significant fraction of the total mass of the object.

The limits presented here for a cutoff in the halo mass function are comparatively higher than the limits for a Warm Dark Matter half-mode mass found using the same lens sample presented in our previous work \citep{JWST_4_Gilman}. In that work,  we found limits of $m_{\rm{hm}}<10^{7.4}$, ($m_{\rm{hm}}<10^{7.2}$) based on the 10:1 odds ratio, using a prior on $\Sigma_{\rm{sub}}$ based on {\tt galacticus} ($N$-body simulations). 

To develop intuition for how these limits compare we can consider the two competing effects. On the one hand, the sharp cutoff model studied here has no field halos at all below the cutoff mass, whereas in WDM models, there are still a significant number of low-mass halos below the half-mode mass. This would tend to mean that the value of $\mlow$ would need to be lower to produce a fixed amount of lensing signal. On the other hand, strong lenses are sensitive to the internal densities of halos \citep[e.g.][]{gilman_constraining_2023, gilman_constraining_2023,minor_unexpected_2021, Despali_concentrations}, with denser halos causing relatively larger perturbations at fixed total halo mass. Because the concentrations of warm dark matter halos are systematically lower in WDM than in CDM, even up to a factor of ten above the half-mode mass, they are less efficient lenses and thus have a lower probability of perturbing a lensed image. Furthermore, the lower concentrations make subhalos more prone to tidal disruption at higher masses, further decreasing the probability that a WDM halo perturbs a lens. The effect of concentrations has the opposite effect to the lower number of small mass halos, it tends to mean that a higher value of $\mlow$ can produce a given lensing signal because the low mass halos are more effective lenses. The fact that the inference on $\mlow$ is weaker than that on $m_{\rm{hm}}$ would suggest that the concentration difference between the CDM and $\mlow$ models is the dominant of these two effects.  

\begin{figure}
    \includegraphics[trim=1cm 1cm 1cm
    0.5cm,width=0.48\textwidth]{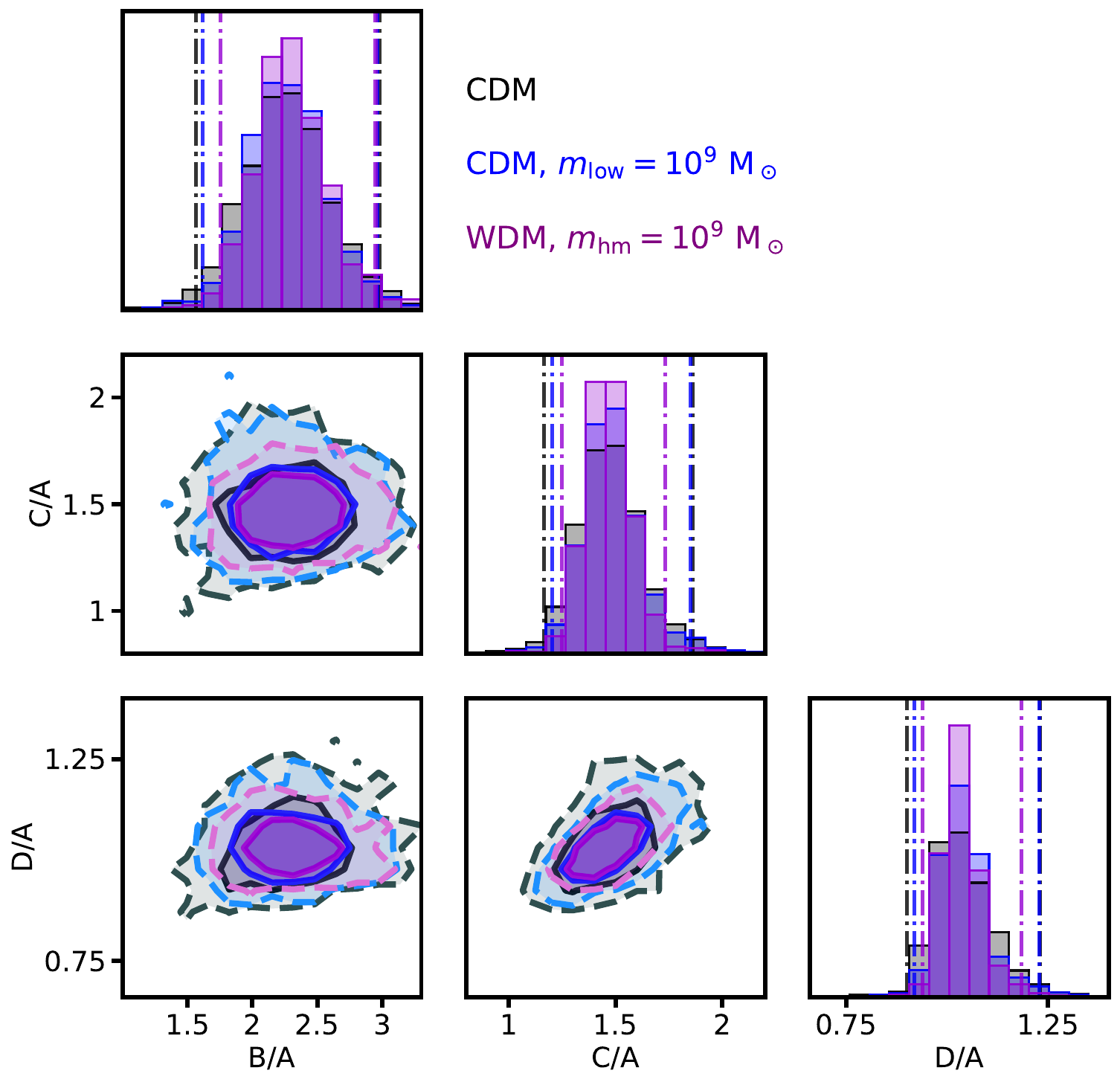}
    \caption{\label{fig:mlow_cdm} Comparison of the flux ratio distribution for a mock cusp lens for three different dark matter scenarios; a warm dark matter model with a half-mode-mass $m_{\rm hm} = 10^9$ M$_\odot$, a CDM-like model with a sharp cutoff at $m_{\rm{low}} = 10^9$ M$_\odot$ and a CDM model with no detectable cutoff. All other dark matter parameters, are held fixed between the simulations. The $\mlow$ model predicts a broader spread of flux ratios than the WDM model, providing a qualitative demonstration of the relative importance of the higher concentrations of $\mlow$ subhalos in perturbing the lensed images. Vertical bars indicate the 95\% confidence interval.}
\end{figure}

We provide a qualitative exploration of these effects in Figure~\ref{fig:mlow_cdm} by showing how the image fluxes for a mock lens are perturbed in a WDM model with $m_{\rm{hm}} = 10^{9}$ M$_\odot$, and a CDM model with a cutoff in the field halo mass function $m_{\rm{low}} = 10^{9}$ M$_\odot$
and finally a CDM model. We note that this example does not map in a trivial way into an expected dark matter constraint for the two models as this would require a measurement of a `true' flux ratio for this system as well as multiplication of probability distributions over many mock systems.
However, the larger spread in the flux ratios for the $\mlow$ model relative to the WDM model demonstrates that the $\mlow$ dark matter halos are more likely to cause a significant perturbation to the measured flux ratios than the WDM halos, highlighting that they are more effective lenses for fixed value of $\mlow$ and $m_{\rm{hm}}$. 
As expected, the additional low mass halos in the true CDM model further broaden the flux ratio distribution relative to the $m_{\rm{low}} = 10^{9}$ M$_\odot$ model.

%Figure with flux ratios. (Possibly in appendix/supplemental information).

%\textit{Discussion}--- The limits presented here are among the strongest limits measured to date on the mass scale at which dark matter halos form in the field. They are comparable to existing limits from Milky Way satellite galaxies by \cite{nadler_milky_2020} who report $\mlow<10^{8.8}, 10^{8.4}$ $M_\odot$ using either only confirmed or confirmed and unconfirmed Milky Way satellites at 95\% confidence. Given that the prior ranges on $\mlow$ used in the two analyses are different, we do not aim to directly compare the stringency of these two measurements, only to demonstrate that both gravitational lensing and Milky Way satellite counts measurements indicate the existence of a significant population of dark matter halos below masses of 10$^{8.5}$ M$_\odot$.

\begin{figure}
    \includegraphics[trim=1cm 1cm 1cm
    0.5cm,width=0.48\textwidth]{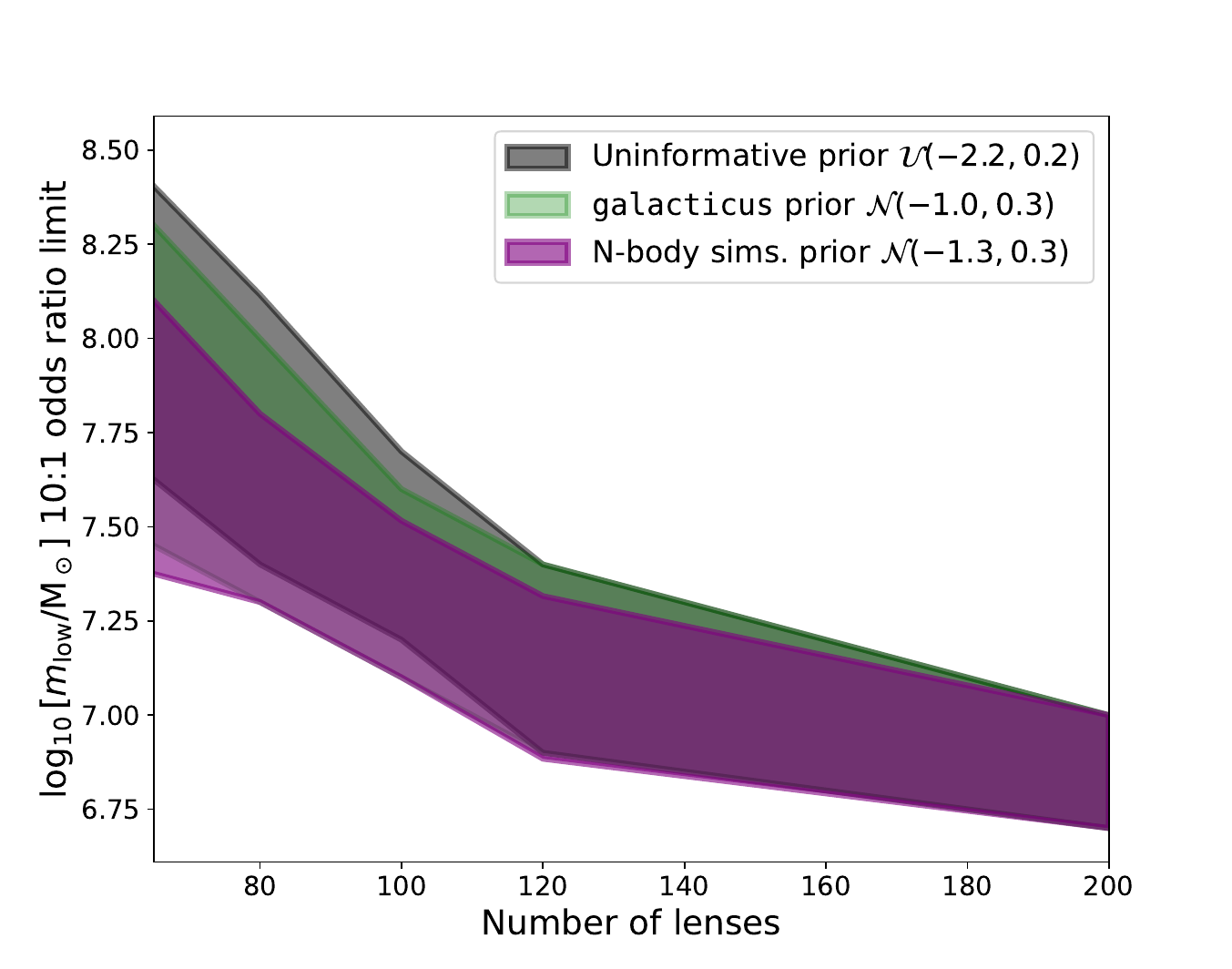}
    \caption{\label{fig:mlow_forecast} 
    Forecast 10:1 posterior odds ratio limit on $\mlow$ as a function of the number of quadruply imaged quasars in the sample.  Spread indicates the 68\% confidence interval from bootstrap resampling while colors represent prior choices on $\Sigma_{\rm{sub}}$. The impact of the informative priors relative to a uniform prior decreases significantly with more than 100 lenses. With about 200 lenses, we forecast a constraint of $\mlow \lesssim 10^{7}$ M$\odot$.
     }
\end{figure}

\emph{Physical Interpretation}--- Physically this measurement provides a conservative limit on particle physics models with a cutoff in the halo mass function. Typically structure formation models predict lower concentrations below the mass cutoff scale for models with a cutoff in the power spectrum, however we show that even in the absence of such a concentration dependence, we can place limits on a cutoff in the halo mass function. 

This measurement also has implications for the galaxy-halo connection at low-mass scales. In particular, it sets the minimum mass scale, assuming a CDM mass-concentration relation, at which dark matter halos must form. A measurement of a lower mass limit that is significantly below the limit measured via methods such as Milky Way satellite counts can confirm for a key prediction of CDM, namely that there exist a significant population of dark matter halos which can only be detected via their gravitational influence. Our current measurement is not able to place a limit at the relevant mass scale for this statement, however, we anticipate significantly improved constraints in the near future.

\emph{Future constraints}--- Thousands of new strong gravitational lenses are forecast to be discovered in the next decade between LSST, Euclid, and Roman \citep{Oguri_forecast_2010}, with $\sim$ 200 quadruply imaged quasars readily observable in the next 5 years based on the expected year one depth of LSST. We use the existing constraints in this work to forecast the projected constraints with additional gravitational lenses. This forecast assumes that the lenses have flux ratio and narrow-line measurements, as well as imaging data with similar precision to those in this study. After lens discovery, flux ratio measurements will require additional follow-up either with JWST for warm-dust flux ratios or ground-based IFU with adaptive optics for narrow-line measurements. Lenses falling within the Roman and Euclid footprints will automatically have imaging quality sufficient for the macromodel constraints presented here, while systems falling outside of these surveys can also be imaged from the ground with adaptive optics.

We simulate future observations by bootstrap resampling from the current individual lens posteriors. Figure~\ref{fig:mlow_forecast} shows the projected constraints for up to 200 lenses. Bootstrapping is performed by drawing randomly from the current lens posterior probability distributions. Contours represent the 68\% confidence interval for 50 randomly drawn samples of lenses. Beyond 120 quadruply imaged quasars, the prior on $\Sigma_{\rm{sub}}$ becomes significantly less important. With 200 lenses, we forecast a constraint on $\mlow$ $\lesssim 10^{7}$ M$_\odot$. 

In addition to new lenses, the combination of this data set with other complementary data sets can greatly strengthen the combined analysis. In particular, \cite{nadler_dark_2021} demonstrated that constraints from Milky Way satellites, combined with strong lensing measurements significantly improved the constraint on the dark matter half mode mass. In the near future the constraint on $\mlow$ can be improved by such a comparison.

%\textit{Forecast for 200 quads}---

%forecast figure

\section*{Acknowledgments}
    %We thank Alex Drlica-Wagner, Josh Frieman, and Ethan Nadler for helpful discussions. 
	
    DG acknowledges support for this work provided by the Brinson Foundation through a Brinson Prize Fellowship grant. 

    AMN and CG acknowledge support from the National Science Foundation through the grant ``CAREER: An order of magnitude improvement in measurements of the physical properties of dark matter" NSF-AST-2442975.
    
    TT, XD and HP acknowledge support from the National Science Foundation through the grant ``Collaborative Research: Measuring the physical properties of dark matter with strong gravitational lensing" NSF-AST-2205100.

    DW acknowledges support by NSF through grants NSF-AST-1906976 and NSF-AST-1836016, and from the Moore Foundation through grant 8548.

    D. Sluse acknowledges the support of the Fonds de la Recherche Scientifique-FNRS, Belgium, under grant No. 4.4503.1 and the Belgian Federal Science Policy Office (BELSPO) for the provision of financial support in the framework of the PRODEX Programme of the European Space Agency (ESA) under contract number 4000142531.

    P.M. acknowledges support from the National Science Foundation through grant NSF-AST-2407277. 

    SB acknowledges support by the Department of Physics and Astronomy, Stony Brook University, and by DoE Grant DE-SC0026113.

    TA acknowledges support from ANID-FONDECYT Regular Project 1240105 and the ANID BASAL project FB210003.

    KNA is partially supported by the U.S. National Science Foundation (NSF) Theoretical Physics Program Grant No.\ PHY-2210283. 

    %V.N.B. acknowledges funding support from STScI grant Nos. HST-GO-17103 and HST-AR-17063. 
    
    SGD acknowledges a generous support from the Ajax Foundation.

    SFH acknowledges support through UK Research and Innovation (UKRI) under the UK government’s Horizon Europe Funding Guarantee (EP/Z533920/1, selected in the 2023 ERC Advanced Grant round) and an STFC Small Award (ST/Y001656/1).
    
    A. K. was supported by the U.S. Department of Energy (DOE) Grant No. DE-SC0009937;  by World Premier International Research Center Initiative (WPI), MEXT, Japan; and by Japan Society for the Promotion of Science (JSPS) KAKENHI Grant No. JP20H05853.
    
    %V.M. acknowledges support from ANID FONDECYT Regular grant number 1231418 and Centro de Astrof\'{\i}sica de Valpara\'{\i}so CIDI 21.

    Part of this work was carried out at the Jet Propulsion Laboratory, California Institute of Technology, under a contract with NASA.

    %MS acknowledges partial support from NASA grant 80NSSC22K1294. 

    K.C.W. is supported by JSPS KAKENHI Grant Numbers JP24K07089, JP24H00221.
	
	This work is based on observations made with the James Webb Space Telescope through the Cycle 1 program JWST GO-2046 (PI:Nierenberg), and the Hubble Space Telescope through HST-GO-15320, HST-GO-15652, HST-GO-17916 (PI:Treu) and HST-GO-13732 (PI:Nierenberg). Funding from NASA through these programs is gratefully acknowledged.

    Some of the data presented herein were obtained at Keck Observatory, which is a private 501(c)3 non-profit organization operated as a scientific partnership among the California Institute of Technology, the University of California, and the National Aeronautics and Space Administration. The Keck facilities we used were LRIS and OSIRIS. The Observatory was made possible by the generous financial support of the W. M. Keck Foundation.  The authors wish to recognize and acknowledge the very significant cultural role and reverence that the summit of Maunakea has always had within the Native Hawaiian community. We are most fortunate to have the opportunity to conduct observations from this mountain.

    This research is based in part on data collected at the Subaru Telescope, which is operated by the National Astronomical Observatory of Japan. We are honored and grateful for the opportunity of observing the Universe from Maunakea, which has cultural, historical, and natural significance in Hawaii.

	This work used computational and storage services provided by the University of Chicago’s Research Computing Center; Caltech's Resnick High Performance Computing Center through Carnegie Science's partnership; the Pinnacles (NSF MRI, $\#$ 2019144) and CENVAL-ARC (NSF $\#$ 2346744) computing clusters at the Cyberinfrastructure and Research Technologies (CIRT) at University of California, Merced; and the Hoffman2 Cluster which is operated by the UCLA Office of Advanced Research Computing’s Research Technology Group.  This research was done using services provided by the OSG Consortium \cite{osg_1, osg_2}, which is supported by the National Science Foundation awards $\#$2030508 and $\#$2323298.
	
	\section*{Software} This work made use of {\tt{astropy}}:\footnote{\url{http://www.astropy.org}} a community-developed core Python package and an ecosystem of tools and resources for astronomy \citep{astropy:2013, astropy:2018, astropy:2022};  {\tt{cobyqa}} \citep{rago_thesis,razh_cobyqa}; {\tt{colossus}} \citep{Diemer18};  {\tt{lenstronomy}}\footnote{\url{https://github.com/lenstronomy/lenstronomy}} \citep{birrer_lenstronomy_2018, birrer_lenstronomy_2021}; {\tt{numpy}} \citep{numpy}; {\tt{pyHalo}}\footnote{\url{https://github.com/dangilman/pyHalo}} \citep{gilman_warm_2020}; {\tt{trikde}}\footnote{\url{https://github.com/dangilman/trikde}}; {\tt{samana}}\footnote{\url{https://github.com/dangilman/samana}}; and {\tt{scipy}} \citep{scipy}. 
	
	\section*{Data availability}
	The data used in this article come from HST-GO-15320, HST-GO-15652, HST-GO-17917, HST-GO-13732 and JWST GO-2046. The raw data are publicly available online. Astrometry and flux ratio measurements are presented by \citet{nierenberg_detection_2014, nierenberg_probing_2017, nierenberg_double_2020,Keeley_2024}. Reduced imaging data for the systems analyzed in this work are available in the open-source software {\tt{samana}}, which also provides notebooks that perform the lens modeling and scripts to reproduce the dark matter analysis.

\bibliography{references} 
%\bibliographystyle{jponew}

%{\bf \large Supplementary Material}
%\input{LensFits}

% Don't change these lines
%\bsp	% typesetting comment
%\label{lastpage}
\end{document}